\documentclass[12pt]{iopart}

\usepackage{epsf}
\usepackage{bm}
\begin{document}

\title{Equation of state at finite baryon density based on lattice 
       QCD}

\author{Pasi Huovinen$^1$ and P\'eter Petreczky$^2$ 
}        

\address{$^1$
Institut f\"ur Theoretische Physik,
Johann Wolfgang Goethe-Universit\"at,\\
60438 Frankfurt am Main, Germany
}

\address{$^2$Physics Department, Brookhaven National Laboratory, 
         Upton, NY 11973, USA}

\ead{huovinen@th.physik.uni-frankfurt.de}

\begin{abstract}
  We employ the lattice QCD data on Taylor expansion coefficients to
  extend our previous parametrization of the equation of state to
  finite baryon density. When we take into account lattice spacing and
  quark mass dependence of the hadron masses, the Taylor coefficients
  at low temperature are equal to those of hadron resonance gas. Thus
  the equation of state is smoothly connected to the hadron resonance
  gas equation of state at low temperatures.  We also show how the
  elliptic flow is affected by this equation of state at the maximum
  SPS energy.\footnote{Talk given at Quark Matter 2011, 22-28 May 2011, 
                       Annecy, France}
\end{abstract}

One of the methods to extend the lattice QCD calculations to non-zero
chemical potential is Taylor expansion. In that approach pressure is
Taylor expanded in chemical potentials, and the Taylor coefficients
are calculated on the lattice at zero chemical potential. In this
contribution we use the results of the most comprehensive lattice QCD
analysis of the Taylor coefficients to date~\cite{Miao:2008,Cheng:2008}
to construct a parametrization of an equation of state (EoS) for
finite baryon density. As in our earlier parametrization of the EoS at
zero chemical potential~\cite{Huovinen:2009}, we require that our
parametrization matches smoothly to the hadron resonance gas (HRG) at
low temperatures.

\begin{figure}[t]
   \hfill
  \epsfysize=60mm \epsfbox{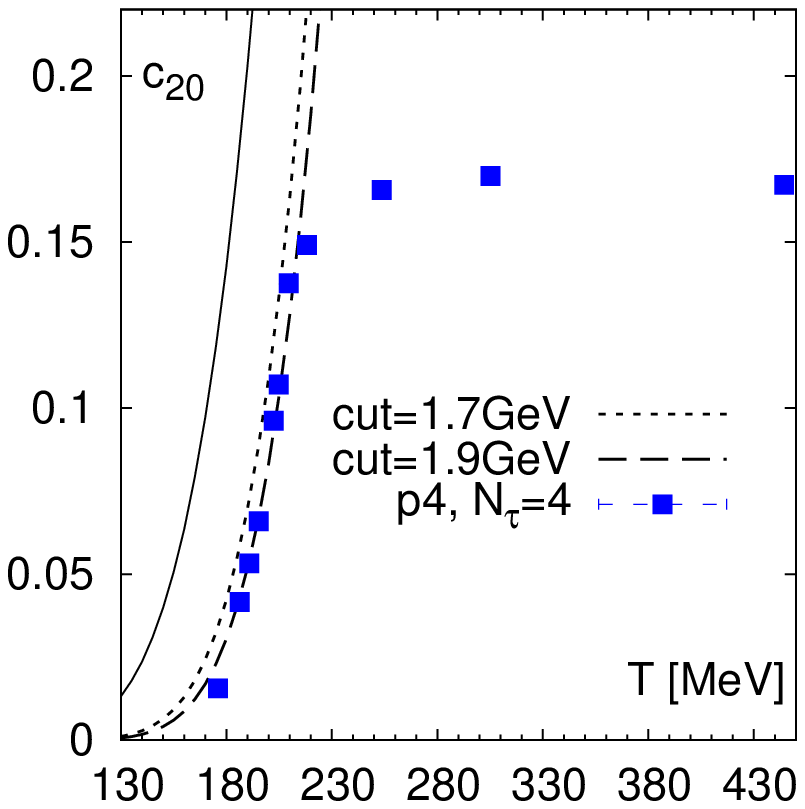}
   \hfill
  \epsfysize=60mm \epsfbox{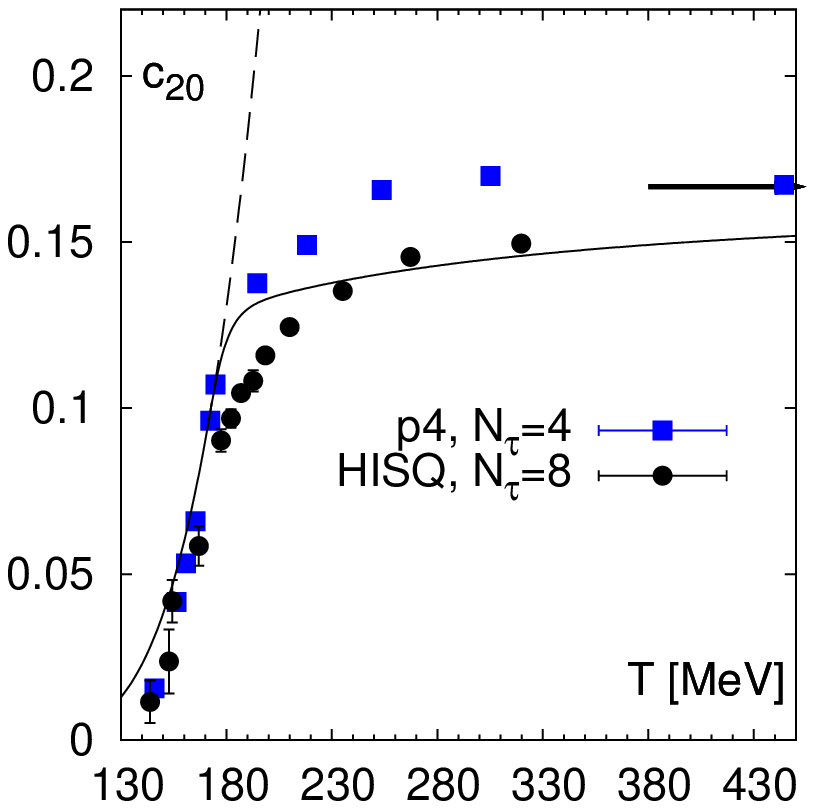}
   \hfill
  \caption{(Left) The second order baryonic Taylor coefficient
    $c_{20}$ calculated on the lattice with p4
    action~\cite{Cheng:2008} and compared with the HRG values with
    free particle (solid line) and lattice masses (dashed and dotted
    lines).\\ (Right) The parametrization (solid line) and HRG value
    (dashed) of the $c_{20}$ coefficient compared with the
    shifted p4 data (see the text). The more recent
    lattice result with the HISQ action~\cite{Bazavov} is also
    shown. The arrow depicts the Stefan-Boltzmann value of $c_{20}$.}
 \label{fig:taylor}
 \end{figure}

Taylor coefficients are simply derivatives of
pressure with respect to baryon and strangeness chemical potential:
\begin{equation}
 c_{ij}(T) = \frac{1}{i!j!}\frac{T^{i+j}}{T^4}\frac{\partial^i}{\partial \mu_B^i}
             \frac{\partial^j}{\partial \mu_S^j}P(T,\mu_B=0,\mu_S=0).
\end{equation}
Purely baryonic and strange coefficients are related to quadratic and
higher order fluctuations of conserved charges, whereas mixed
derivatives of pressure give correlations of these charges.  The
second order baryonic coefficient $c_{20}$ is shown in the left panel
of Fig.~\ref{fig:taylor}. The lattice result for $c_{20}$, as well as
for all the other coefficients, is well below the HRG result obtained
with physical masses (solid line). This discrepancy can largely be
explained by the lattice discretization effects on hadron masses: When
the hadron mass spectrum is modified accordingly (for details
see~\cite{Huovinen:2010}), the HRG model reproduces the lattice
data. The dashed and dotted curves in Fig.~\ref{fig:taylor} refer to
different treatment of baryonic resonances. We modify their masses in
the same way than ground state baryons up to a threshold $m_{cut}$ but
keep the masses of heavier resonances in their physical values. As
seen, the exact value of this threshold has only a small effect on
$c_{20}$.

 Closer look at the HRG curves in Fig.~\ref{fig:taylor} (left) reveals
 that the incorporation of lattice masses has basically shifted the
 HRG curve by 30 MeV towards higher temperatures. This can be clearly
 seen in the right panel of Fig.~\ref{fig:taylor} where we plot the HRG
 curve with physical masses (dashed line) and compare it with the
 lattice data, where we have shifted all the points below 206 MeV
 temperature by 30 MeV, and the 209 MeV point by 15 MeV, towards lower
 temperature. Now the data points which agreed well with the HRG curve
 with lattice masses, agree well with the HRG curve with physical
 masses. Thus we propose that deviation from the continuum limit for
 the p4 data at low temperatures can be accounted for by shifting the
 data points by 30 MeV to lower temperature. For further confirmation
 of this procedure we also plot the recent HISQ result of
 $c_{20}$~\cite{Bazavov} in Fig.~\ref{fig:taylor} (right): At low
 temperatures the shifted p4 data agree with the HISQ data.  In the
 strange sector the discretization effects are slightly smaller than
 shown in Fig.~\ref{fig:taylor}. However, for simplicity we use the
 same shift of 30 MeV for all the coefficients.

 We parametrize the shifted data using an inverse polynomial of three
 ($c_{20}$), four ($c_{11}$ and $c_{02}$), or five (fourth and sixth
 order coefficients) terms:
\begin{equation}
   c_{ij}(T) = \frac{a_{1ij}}{T^{n_{1ij}}} + \frac{a_{2ij}}{T^{n_{2ij}}}
             + \frac{a_{3ij}}{T^{n_{3ij}}} + \frac{a_{4ij}}{T^{n_{4ij}}}
             + \frac{a_{5ij}}{T^{n_{5ij}}} + c_{ij}^\mathrm{SB},
\end{equation}
where $c_{ij}^\mathrm{SB}$ is the Stefan-Boltzmann value of the
particular coefficient, and the powers $n_{kij}$ are required to be
integers between 1 and 42. As in our parametrization of the EoS at
zero net baryon density~\cite{Huovinen:2009}, we match this
parametrization to the HRG value at temperature $T_\mathrm{SW}$ by
requiring that the Taylor coefficient and its first and second
derivatives are continuous. Since the recent lattice data obtained
using HISQ action~\cite{Bazavov} shows that the second order
coefficients approach their Stefan-Boltzmann limits slowly, we require
that their value is 95\% of their Stefan-Boltzmann value at 800 MeV
temperature. These constraints fix three (or four) of the parameters
$a_{kij}$. The remaining parameters, including the switching
temperatures, are fixed by a $\chi^2$ fit to the lattice data. As an
example we show the parametrized $c_{20}$ in the right panel of
Fig.~\ref{fig:taylor}.

 \begin{figure}[thb]
  \epsfysize=60mm \epsfbox{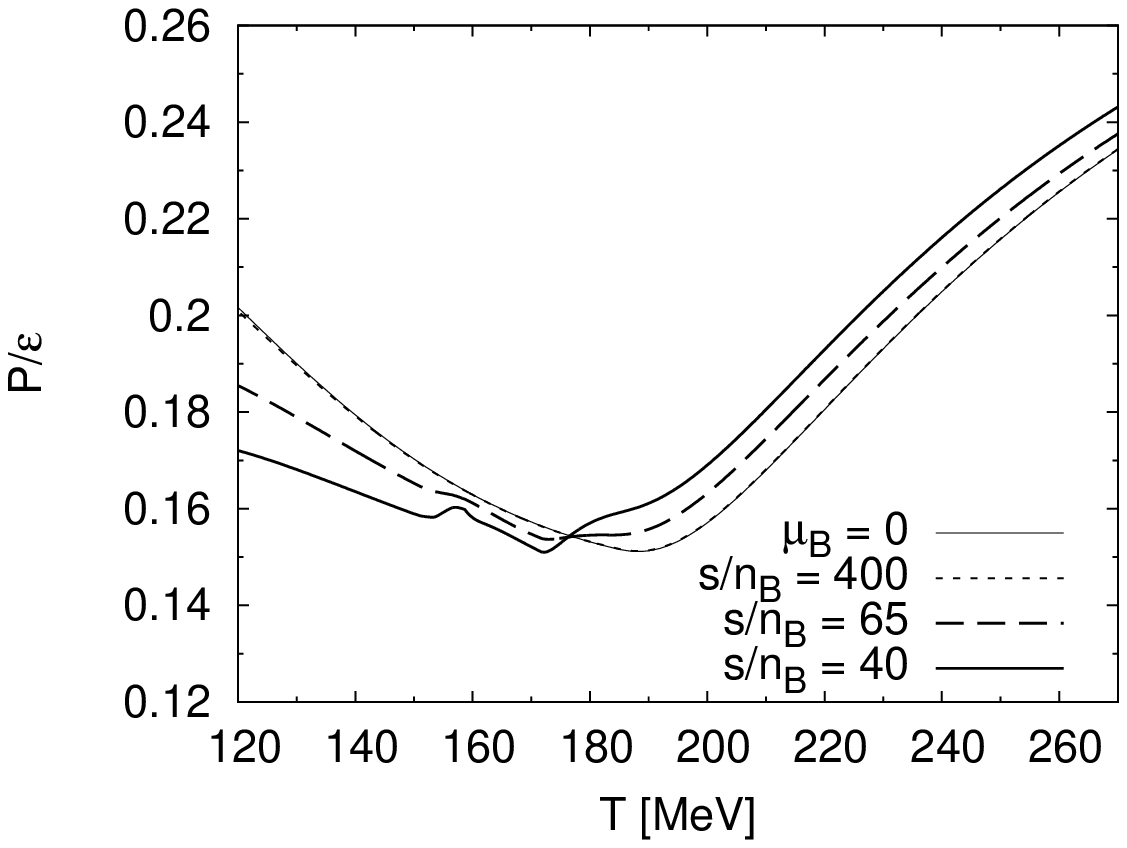}
   \hfill
  \epsfysize=60mm \epsfbox{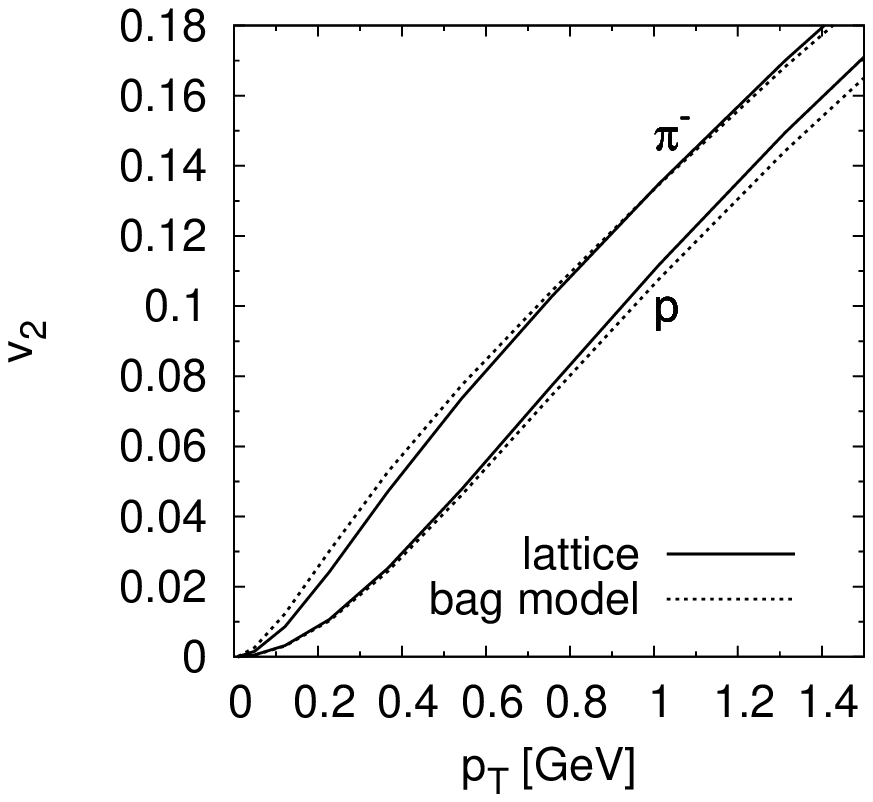}
  \caption{(Left) Pressure over energy density as function of
    temperature on various isentropic curves with constant entropy per
    baryon.\\ (Right) $p_T$-differential elliptic flow of pions (upper
    curves) and protons (lower curves) in an ideal fluid simulation of
    $\sqrt{s_\mathrm{NN}} = 17$ GeV Pb+Pb collisions at $b=7$ fm.}
 \label{fig:eos}
 \end{figure}

Once the coefficients are known, pressure can be written as
\begin{equation}
\frac{P}{T^4} = \sum_{ij} c_{ij}(T) \left(\frac{\mu_B}{T}\right)^i
                                  \left(\frac{\mu_S}{T}\right)^j,
 \label{PT4}
\end{equation}
and all the other thermodynamical quantities can be obtained from
Eq.(\ref{PT4}) by using the laws of thermodynamics. As pressure at
$\mu_B=0$, i.e.\ the coefficient $c_{00}$, we use our earlier
parametrization $s95p$-v1~\cite{Huovinen:2009}.  We describe the EoS
in the left panel of Fig.~\ref{fig:eos} by showing the pressure to
energy density ratio on various isentropic curves with constant
entropy per baryon. The curves at $s/n_B=400$, 65, and 40 are relevant
at collision energies $\sqrt{s_\mathrm{NN}}=200$, 39 and 17 GeV,
respectively. At $s/n_B=400$ (dotted line), the EoS is basically
identical to the EoS at $\mu_B = 0$ (thin solid line). This vindicates
the common approximation of ignoring the finite net baryon density in
the description of collisions at the full RHIC energy
($\sqrt{s_\mathrm{NN}}=200$ GeV). At larger baryon densities the
effect of finite baryon density is no longer negligible. The larger
the density, the stiffer the EoS above, and softer below the
transition temperature. Furthermore, additional structure begins to
appear around the transition temperature with increasing density.
This structure is mostly an unphysical artefact of our fitting
procedure. We required the two first derivatives with respect to
temperature to be continuous, but the speed of sound is proportional
to the second derivative of the coefficients. Thus, in our
parametrization, the derivative of the speed of sound is not
continuous, and ripples may appear at a switching temperature of any
coefficient. Nevertheless, when pressure is plotted as a function of
energy density, these structures are hardly visible. Therefore we do
not expect them to affect the buildup of flow and the evolution of the
system, and consider our parametrization a reasonable first
attempt. The work to obtain a smoother and better constrained
parametrization is in progress.

We illustrate the effect of the EoS on flow by studying elliptic flow
in Pb+Pb collision at the full SPS collision energy
($\sqrt{s_\mathrm{NN}}=17$ GeV). For simplicity we use a boost
invariant ideal hydrodynamical model to compare our lattice based EoS
to a bag model EoS with a first order phase transition. We tune the
calculation to reproduce the NA49 data for negative hadrons and net
protons in the most central collisions~\cite{NA49}, and use freeze-out
temperatures of $T_\mathrm{dec} = 130$ and 120 MeV for the lattice and
bag model EoSs, respectively.  Since it is very difficult to reproduce
the elliptic flow data at SPS using ideal hydrodynamics, we do not try
to fit the data. Instead we calculate the $p_T$-differential $v_2$ of
pions and protons at fixed impact parameter of $b=7$ fm, see the right
panel of Fig.~\ref{fig:eos}. At RHIC, the pion $v_2(p_T)$ is
insensitive to the EoS, but the proton $v_2(p_T)$ shows a clear
dependence on it~\cite{Huovinen:2005}. However, at lower collision energy
the behaviour is different: Proton $v_2(p_T)$ is as insensitive to the
EoS as the pion $v_2(p_T)$. This behaviour is supported by the early
ideal fluid calculations of $v_2$: 
It was seen that at SPS both a bag model EoS and a purely hadronic EoS
led to quite a similar $v_2(p_T)$~\cite{Kolb:2000fha}, but at RHIC a
purely hadronic and lattice EoS led to a similar proton $v_2(p_T)$,
whereas the bag model EoS lead to a smaller proton
$v_2(p_T)$~\cite{Huovinen:2005}.

To summarise, we have shown that a temperature shift of 30 MeV is a
good approximation of the discretization effects in the lattice QCD
data obtained using p4 action. We have constructed an equation of
state for finite baryon densities based on hadron resonance gas and
lattice QCD data. At the full SPS energy ($\sqrt{s_\mathrm{NN}} = 17$
GeV) the $p_T$-differential elliptic flow is almost insensitive to the
equation of state. This is bad news for the experimental search of the
critical point, since a change from a first order phase transition to
a smooth crossover does not cause an observable change in the flow.

\ack 
This work was supported by BMBF under contract no.\ 06FY9092, and by
the U.S. Department of Energy under contract DE-AC02-98CH1086.

\end{document}